\documentclass{ws-procs9x6}

\newcommand{\be}{\begin{equation}}
\newcommand{\ee}{\end{equation}}
\newcommand{\bea}{\begin{eqnarray}}
\newcommand{\eea}{\end{eqnarray}}

\newcommand{\bfk}{\mbox{\boldmath $k$}}

\newcommand{\pup}{p^\uparrow}

\newcommand{\Adown}{A^\downarrow}

\newcommand{\Aup}{A^\uparrow}

\newcommand{\la}{\lambda}
\def\lsim{\mathrel{\rlap{\lower4pt\hbox{\hskip1pt$\sim$}}\raise1pt\hbox{$<$}}}
\def\gsim{\mathrel{\rlap{\lower4pt\hbox{\hskip1pt$\sim$}}\raise1pt\hbox{$>$}}}
\def\nostrocostruttino#1\over#2{\mathrel{\mathop{\kern 0pt \rlap
{\hbox{$#1$}}} \hbox{\kern-.135em $#2$}}}
\def\sumint{\nostrocostruttino \sum \over {\displaystyle\int}}
\newcommand{\cal}{\mathcal}

\newcommand{\NP}[1]{{\it Nucl.\ Phys.}\ {\bf #1}}

\newcommand{\PL}[1]{{\it Phys.\ Lett.}\ {\bf #1}}
\newcommand{\PR}[1]{{\it Phys.\ Rev.}\ {\bf #1}}
\newcommand{\PRL}[1]{{\it Phys.\ Rev.\ Lett.}\ {\bf #1}}

\begin{document}

\title{Can the Collins mechanism explain \\
the large transverse single spin asymmetries \\
observed 
in $\pup\! \lowercase{ p} \to \pi\, X$?\footnote{\uppercase{T}alk delivered by 
\uppercase{U}.~\uppercase{D}'\uppercase{A}lesio at the
``16th \uppercase{I}nternational \uppercase{S}pin \uppercase{P}hysics 
\uppercase{S}ymposium'', \uppercase{SPIN}2004, 
\uppercase{O}ctober 10-16, 2004, \uppercase{T}rieste, \uppercase{I}taly. }
}

\author{M. ANSELMINO,$^1$ M. BOGLIONE,$^1$ U. D'ALESIO,$^2$\\
  E. LEADER,$^3$ F. MURGIA$^2$}
 
\address{
$^1$Dipartimento di Fisica Teorica, Universit\`a di Torino and \\
          INFN, Sezione di Torino, Via P. Giuria 1, I-10125 Torino, Italy\\
$^2$Dipartimento di Fisica, Universit\`a di Cagliari and\\
INFN, Sezione di Cagliari, C.P. 170, I-09042 Monserrato (CA), Italy\\
$^3$Imperial College, Prince Consort Road, London, SW7 2BW, England}

\maketitle

\abstracts{
We present a calculation
of inclusive polarised and
unpolarised cross sections within pQCD and the factorisation scheme, taking
into account the parton intrinsic motion, $\bfk_\perp$,  
in distribution and fragmentation
functions, as well as in the elementary dynamics. 
We show, in contradiction with
earlier claims, that the Collins mechanism is suppressed and 
unable to explain the large
asymmetries found in $\pup \, p \to \pi \, X$ at moderate to large Feynman
$x_F$. The Sivers effect is not suppressed.
}

In the standard perturbative QCD approach to
inclusive particle production at high
energies, intrinsic transverse motions 
are integrated out. 
Nevertheless, we know how  they can help in describing 
experimental data for inclusive particle production in 
hadronic processes at moderately
large $p_T$,\cite{dm04} otherwise heavily underestimated. 

When we consider polarised cross sections, $\bfk_\perp$ could become
essential: certain spin and $\bfk$-dependent
effects, generated by soft mechanisms, can be used to understand 
the large transverse single spin asymmetries (SSA) found in 
many reactions like $p^{\uparrow} p \to \pi  X$.

For polarised processes, $(A,S_A) + (B,S_B) \to C + X$, 
by introducing in the factorisation scheme,
in addition to the distribution functions, the helicity density
matrices which describe the parton spin states, we have 
\begin{eqnarray}
\label{gen1}
d\sigma^{(A,S_A) + (B,S_B) \to C + X}= 
\sum_{a,b,c,d, \{\la\}} \rho_{\la^{\,}_a,\la^{\prime}_a}^{a/A,S_A} \, 
\hat f_{a/A,S_A}(x_a,\bfk_{\perp a})\qquad
&& \\
\otimes\,\rho_{\la^{\,}_b, \la^{\prime}_b}^{b/B,S_B} \,
\hat f_{b/B,S_B}(x_b,\bfk_{\perp b})\otimes 
\hat M_{\la^{\,}_c, \la^{\,}_d; \la^{\,}_a, \la^{\,}_b} \,
\hat M^*_{\la^{\prime}_c, \la^{\,}_d; \la^{\prime}_a,\la^{\prime}_b}
\otimes 
\hat{D}^{\la^{\,}_C,\la^{\,}_C}_{\la^{\,}_c,\la^{\prime}_c}(z,\bfk_{\perp
  C}), &&\nonumber
\end{eqnarray}
where $\{\la\}$ is a shorthand for all helicity indices involved.
In Eq.~(\ref{gen1}), 
the $\hat M_{\la^{\,}_c, \la^{\,}_d; \la^{\,}_a, \la^{\,}_b}$'s 
are the helicity amplitudes for the hard process $ab\to cd$  
and $\hat D^{\la^{\,}_C,\la^{\prime}_C}_{\la^{\,}_c,\la^{\prime}_c}(z,
\bfk_{\perp C})$ is the product of {\it fragmentation amplitudes} for the
$c \to C + X$ process. The helicity density matrix,  
$\rho_{\la^{\,}_a,\la^{\prime}_a}^{a/A,S_A}$,  
of parton $a$ inside hadron $A$ with polarisation $S_{A}$,  
is identically equal to
\bea
\rho_{\la^{\,}_a, \la^{\prime}_a}^{a/A,S_A} \>
\hat f_{a/A,S_A}(x_a,\bfk_{\perp a})
&=& \sum_{\la^{\,}_A, \la^{\prime}_A}
\rho_{\la^{\,}_A, \la^{\prime}_A}^{A,S_A}
\sumint_{X_A, \la_{X_A}} \!\!\!\!\!
{\hat{\cal F}}_{\la^{\,}_a, \la^{\,}_{X_A};
\la^{\,}_A} \, {\hat{\cal F}}^*_{\la^{\prime}_a,\la^{\,}_{X_A}; \la^{\prime}_A}
\label{distramp} \nonumber\\
&=& \sum_{\la^{\,}_A, \la^{\prime}_A}
\rho_{\la^{\,}_A, \la^{\prime}_A}^{A,S_A} \>
\hat F_{\la^{\,}_A, \la^{\prime}_A}^{\la^{\,}_a,\la^{\prime}_a} \>,\label{defF}
\eea
where $\sumint_{X_A, \la_{X_A}}\!\!\!$ stands for a spin sum and phase
space integration over all undetected remnants of hadron $A$, considered
as a system $X_A$ and the $\hat{\cal F}$'s are the {\it helicity distribution 
amplitudes} for the $A \to a + X$ process.

By using Eq. (\ref{defF}), Eq. (\ref{gen1}) becomes 
\begin{eqnarray}
&& d\sigma^{(A,S_A) + (B,S_B) \to C + X} =  
\sum_{a,b,c,d, \{\la\}}\rho_{\la^{\,}_A, \la^{\prime}_A}^{A,S_A}
\>\hat F_{\la^{\,}_A, \la^{\prime}_A}^{\la^{\,}_a,\la^{\prime}_a}
\otimes \rho_{\la^{\,}_B, \la^{\prime}_B}^{B,S_B}
 \> \hat{F}_{\la^{\,}_B, \la^{\prime}_B}^{\la^{\,}_b,\la^{\prime}_b}
\nonumber \\
&&\otimes\;\hat M_{\la^{\,}_c, \la^{\,}_d; \la^{\,}_a, \la^{\,}_b}
\,\hat M^*_{\la^{\prime}_c, \la^{\,}_d;\la^{\prime}_a,\la^{\prime}_b}
\,
\otimes
\hat{D}^{\la^{\,}_C,\la^{\,}_C}_{\la^{\,}_c,\la^{\prime}_c}
\equiv \sum_{a,b,c,d} \int\! d[PS]\; \Sigma(S_A,S_B)
\,.\label{gen3}
\end{eqnarray}
Eq.~(\ref{gen3}) contains all possible combinations of different
distribution and fragmentation amplitudes, with definite
partonic interpretations.

Let us now study the processes $\Aup(\Adown) \, B \to \pi \, X$
in the $AB$ center of mass frame, with the polarised beam moving along
the positive $Z$-axis and the pion produced in the $XZ$-plane. 
The $\uparrow (\downarrow)$ is the $+Y(-Y)$ direction. 
Here we focus only on the contribution of the 
Collins mechanism,\cite{col} that is the azimuthal dependence of the 
number of pions created in the fragmentation of a transversely polarised
quark.

The helicity amplitudes $\hat{M}$
in Eq.~(\ref{gen3}), defined in the hadronic c.m.~frame, 
can be related to  those given in the canonical
partonic c.m.~frame, $\hat{M}^{0}$, with $Z$ in the direction of the colliding 
partons and the $XZ$-plane as the scattering plane. 
By performing  a sequence of boost and rotations we get 
\bea
&&\hat M_{\la^{\,}_c, \la^{\,}_d; \la^{\,}_a, \la^{\,}_b} =
 \hat M^0_{\la^{\,}_c, \la^{\,}_d; \la^{\,}_a, \la^{\,}_b} \\
&&\qquad\qquad\times e^{-i (\la^{\,}_a \xi _a + \la^{\,}_b \xi _b -
          \la^{\,}_c \xi _c - \la^{\,}_d \xi _d)}
\,  e^{-i [(\la^{\,}_a - \la^{\,}_b) \tilde \xi _a -
         (\la^{\,}_c - \la^{\,}_d) \tilde \xi _c]}
\, e^{i(\la^{\,}_a - \la^{\,}_b)\phi^{\prime\prime}_c}\,,\nonumber
\label{M-M0}
\eea
where $\xi_i$, $\tilde{\xi}_{i}$ ($i=a,b,c,d$) and  $\phi^{\prime\prime}_c$ 
depend on parton momenta.\cite{N1}

On summing over $\{\la\}$ [Eq. (\ref{gen3})]  
we obtain for the Collins contribution 
to the numerator of SSA ($qb\to qb,\, b=q,\bar q,g$):
\bea
\hskip -.7truecm
\left[ \Sigma(\uparrow,0) - \Sigma(\downarrow,0) \right] &=&
\Bigl\{ F_{+-}^{+-}(x_a, k_{\perp a}) \,
\cos [\phi_a + \phi_c^{\prime\prime} - \xi _a  - \tilde \xi _a
+ \xi _c + \tilde \xi _c + \phi_\pi^H] \nonumber \\
&-& \>\> F_{-+}^{+-}(x_a, k_{\perp a}) \, \cos [\phi_a -
\phi_c^{\prime\prime} + \xi _a  + \tilde \xi _a - \xi _c - \tilde
\xi _c - \phi_\pi^H] \Bigr\}
\label{numan} \nonumber \\
&& \hskip -.35truecm \times \, \hat f_{b/B}(x_b, k_{\perp b}) \>
\hat M_{+,+;+,+}^0 \hat M_{-,+;-,+}^0 \> \left[ -2i
D_{+-}^\pi(z, k_{\perp \pi}) \right] 
\eea
where $\pm = \pm 1/2$ (quarks), $\pm 1$ (gluons), and
$\phi_\pi^H$ is the azimuthal angle of the pion
momentum in the fragmenting quark helicity frame. 
In the notations of Refs. [\refcite{pff1}] (details will be given in
[\refcite{noi}], see also [\refcite{trento}]), we have 
\bea
F_{+-}^{+-}(x, k_\perp) &=& h_1(x, k_\perp) = h_{1T}(x, k_\perp) + 
\frac{k_\perp^2}{2M_p^2} \, h_{1T}^{\perp}(x, k_\perp) \label{fmul1} \\  
F^{+-}_{-+}(x, k_\perp) &=& \frac{k_\perp^2}{2M_p^2} \, 
h_{1T}^{\perp}(x, k_\perp) \label{fmul2} \\ 
-2iD^\pi_{+-}(z, k_\perp) &=& \Delta^ND_{\pi/q^{\uparrow}}(z, k_\perp) = 
\frac{2k_\perp}{zM_\pi} \, H_1^{\perp q}(z, k_\perp) \>, \label{dmul}
\eea
where $M_p$ and $M_\pi$ are respectively the proton and pion mass.
 
In all previous studies the large SSA found 
in the E704 experiment\cite{e704} was explained  
by either the Sivers\cite{noi1} or the Collins mechanisms.\cite{noi2} 
However, only a simplified kinematics was adopted. 
We now believe that the phases involved, when the
kinematics is treated carefully, are crucial, and 
lead to a large suppression of the asymmetry due to the Collins
mechanism. 
Almost no suppression of $A_N$ from the Sivers mechanism is found.\cite{dm04}

In order to demonstrate the extent of the suppression we choose 
for the unmeasured soft functions in Eq.~(\ref{numan}) their known 
upper bounds\cite{bbhm} and adjust their signs so that the contributions 
from the valence flavours (up and down) reinforce each other. 
The results are presented in Fig.~1, which shows 
$(A_N)^{\rm Collins}_{\rm max}$ as a function of $x_F$, at $p_T = 1.5$ 
GeV/$c$ and $\sqrt s \simeq 19.4$ GeV together with the E704 data.\cite{e704} 
The only difference between 
the plots is given by different choices of the polarised distribution 
functions and/or the unpolarised fragmentation functions. 
In the upper-right plot the curves for charged pions obtained by setting
all phases in Eq.~(\ref{numan}) to zero are also given (thin lines). 
We can therefore conclude that the Collins mechanism alone, even 
maximising all its effects, cannot explain the observed SSA values. 

An equally important consequence 
of keeping proper phases emerges in the calculation of SSA at
 $x_F<0$: even maximising all contributions (Sivers mechanism too) 
one gets much smaller (a few \%) $A_N$ values.\cite{noi}

Once more, the importance and subtleties of spin 
effects come out; 
all phases, properly considered, often play 
crucial and unexpected roles. 
\begin{figure} [h,t]
\centering
\includegraphics[width=.7\textwidth, angle=-90]{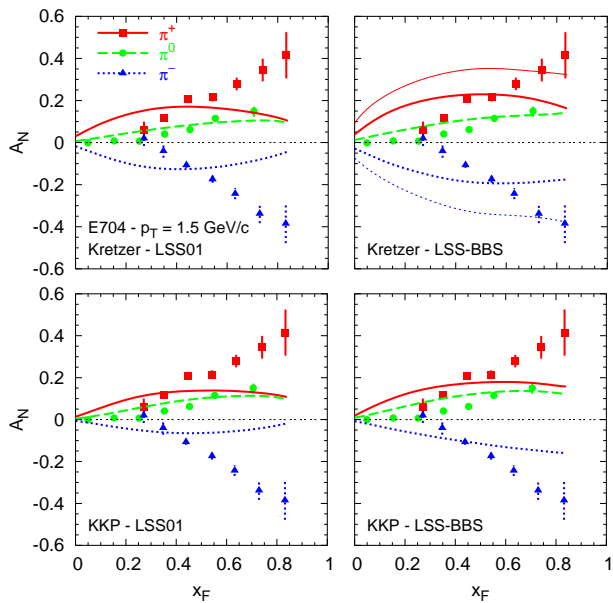}
\caption{Maximised values of $A_N$ vs.~$x_F$, 
as given by the Collins mechanism alone. 
Data are from Ref. [\protect\refcite{e704}]. See the text for further
details.
}
\label{figs}
\end{figure}

This research is part of the 
``EU Integrated Infrastructure Initiative Hadron Physics''
project under contrat number RII3-CT-2004-506078. 
U.D. and F.M. acknowledge partial 
support from ``Cofinanziamento MURST-PRIN03''.

\end{document}